\documentclass[aip,apl,reprint]{revtex4-1}
\usepackage{graphicx}
\usepackage{dcolumn}
\usepackage{bm}
\newcommand{\refer}[2]{~\ref{fig:#1}{(#2)}}

\draft 

\begin{document}


\title{Studying DNA translocation in nanocapillaries using single molecule fluorescence} 



\author{Vivek V. Thacker}
\affiliation{Cavendish Laboratory, University of Cambridge, JJ Thompson Avenue, CB3 0HE, UK}

\author{Sandip Ghosal}
\altaffiliation[Permanent address: ]{Department of Mechanical Engineering, Northwestern University, 2145 Sheridan Road, Evanston, Illinois 60208, USA}
\affiliation{Cavendish Laboratory, University of Cambridge, JJ Thompson Avenue, CB3 0HE, UK}

\author{Silvia Hern\'{a}ndez-Ainsa}
\affiliation{Cavendish Laboratory, University of Cambridge, JJ Thompson Avenue, CB3 0HE, UK}

\author{Nicholas A. W. Bell}
\affiliation{Cavendish Laboratory, University of Cambridge, JJ Thompson Avenue, CB3 0HE, UK}

\author{Ulrich F. Keyser}
\affiliation{Cavendish Laboratory, University of Cambridge, JJ Thompson Avenue, CB3 0HE, UK}

\date{\today}

\begin{abstract}
We demonstrate simultaneous measurements of DNA translocation into glass nanopores using ionic current detection and fluorescent imaging. We verify the correspondence between the passage of a single DNA molecule through the nanopore and the accompanying characteristic ionic current blockage. By tracking the motion of individual DNA molecules in the nanocapillary perpendicular to the optical axis and using a model, we can extract an effective mobility constant for DNA in our geometry under high electric fields.
\end{abstract}


\maketitle 

Since the seminal work on DNA translocations by Kasianowicz \textit{et al.}\cite{Kasianowicz1996a}, there has been an explosion of interest in nanopore based DNA sensing using both solid state and biological nanopores. The basis of nanopore detection is the resistive pulse technique, whereby the presence of a DNA molecule within the pore reduces the flow of ions through the pore. Recent advances in the field such as the devlopment of hybrid\cite{Hall2010a}, DNA origami\cite{Bell2012,Wei2012a} and graphene nanopores\cite{Schneider2010,Merchant2010}, lipid based coating\cite{Yusko2011a, Hernandez-Ainsa2012} as well as the combination with other single molecule techniques such as optical tweezers\cite{Keyser2006a} have greatly enhanced the sensing capabilities of nanopores. 

Where nanopores have been combined with single molecule fluorescence, this has been done with the aim of developing alternative DNA sequencing techniques\cite{McNally2010a}. Previous work has also used single molecule fluorescence for hydrodynamic studies in nanochannels\cite{Stein2010}. In this paper, we demonstrate simultaneous ionic current and fluorescent detection of DNA translocation through glass nanocapillaries.  These have been shown to be an alternative to traditional solid state nanopores\cite{Steinbock2010}. We have extended this system by combining ionic current detection with single molecule fluorescence imaging and have shown by direct visual observation that the drop in ionic current does in fact correspond to the transient blockade of the pore by the DNA molecule. By tracking the movement of the $\lambda$ DNA molecules into the nanocapillary we are also able to extract an effective mobility constant for long DNA molecules.

 Figure\refer{figure1}{a} shows a schematic of the custom built setup for combined fluorescence and ionic current detection. As described previously\cite{Steinbock2010}, a nanocapillary is produced from quartz capillaries of outer diameter 0.5 mm and inner diameter 0.2 mm (Sutter) using a commercial laser puller (Sutter, P2000). This is assembled into a PDMS chip where it connects two reservoirs filled with an electrolyte solution. For the experiments described in this paper we use 100 mM KCl in 100 mM citrate buffer at pH 4.6. The Ag/AgCl electrode within the nanocapillary is held at a constant positive potential with reference to the ground electrode in the reservoir surrounding the nanocapillary. An amplifier (HEKA, EPC 800)  detects the ionic current flowing through the nanocapillary at a bandwidth of 5 kHz. Negatively charged DNA translocates into the nanocapillary which results in characteristic blockades in the ionic current. The I-V curve for the nanocapillary (diameter $\sim$ 50 nm) used for all experiments described in this paper is shown in Figure\refer{figure1}{a}. 

\begin{figure}[h!]
\centering
\includegraphics {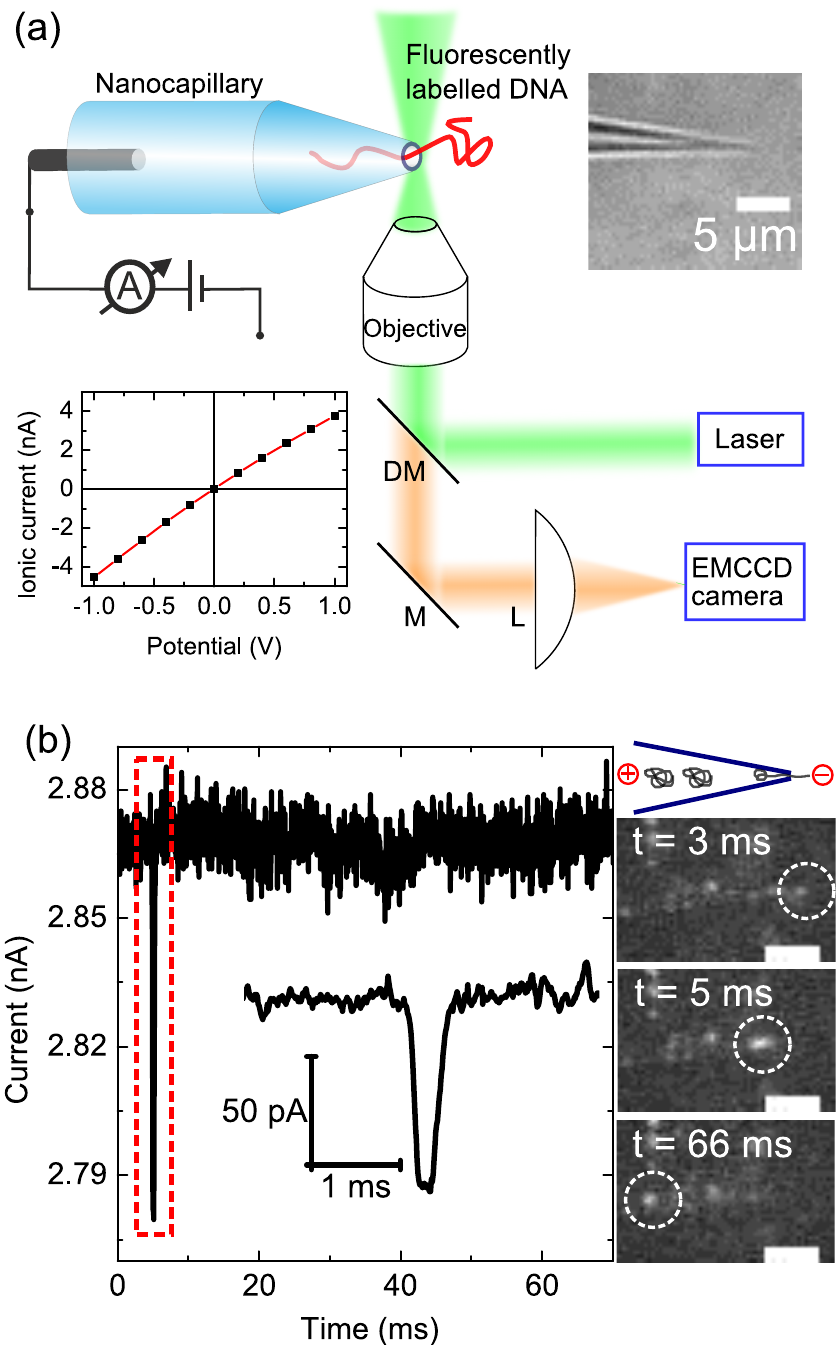}
 \caption{\label{fig:figure1}(a) A setup for combined ionic current detection and fluorescent imaging. Optical components include DM (dichroic mirror), M (mirror) and L (lens). (b) The $\lambda$ DNA molecule is indicated by the circle and translocates (right to left) into the nanocapillary as indicated in the schematic (top, right). The scale bar in the images corresponds to 5 $\mu$m. The ionic current signal of the corresponding event is shown for the time period covered by the three images. A close- up of the translocation event is also presented as an inset.}
\end{figure}

The $\lambda$ DNA stock solution (Fermentas, 0.3 mg/ml, 48502 bp) is diluted to 10 pM in the buffer solution and incubated with 50 nM solution of the DNA intercalating dye SYTOX Orange (Invitrogen), following a protocol described in Yan \textit{et al.}\cite{Yan2000} to obtain a dye:DNA base pair ratio of $\sim$1:10. The dye molecules are excited by emissions from a 532 nm laser (Laser Quantum) operating at $<$ 5mW. A long pass dichroic mirror at 532 nm (Semrock Filters) directs the excitation light to the rear of a 60x objective (UPLSAPO NA=1.2, Olympus). The fluorescence emission from the dye molecules is collected via the same objective. As it is red shifted from the excitation light, it passes through the dichroic mirror and is focused via a mirror and a tube lens to the sensor of a fast electron multipying CCD (EMCCD) camera (Andor iXON3 860). We are able to perform imaging of the translocation processes at a maximum frame rate of 510.2 fps. A bright field image of the nanocapillary used is shown as a top inset in Figure\refer{figure1}{a} and corresponds to images shown in Figure\refer{figure1}{b}. Data acquisition for both ionic current and fluorescence images is done through custom software written in LabVIEW and a DAQ card (PCIe-6351, National Instruments). We use low salt concentrations to reduce non-specific interactions between the dye molecules and the surface of the nanocapillaries. At pH=4.6, the surface charge on the glass is reduced and this reduces the electroosmotic flow in the nanocapillary.

A typical result is shown in Figure\refer{figure1}{b}. $\lambda$ DNA molecules are attracted into the nanocapillary by applying a potential difference of 800 mV. Three images from a video are shown that correspond directly to the single $\lambda$ DNA translocation detected by the resistive pulse technique (indicated by a red box). We set $t=0$ about 5 ms before the the ionic current event. In the first image ($t=3$ ms) the DNA molecule under consideration is freely diffusing in the reservoir in front of the nanocapillary. The second image ($t=5$ ms) shows the start of translocation into the nanocapillary which is accompanied by the ionic current blockade. Finally, we are able to track the DNA molecule upstream into the nanocapillary as shown in the third image ($t=66$ ms). See supplementary material at [URL will be inserted by AIP] for a video of several translocation events and the corresponding ionic current trace. Our results unambiguously verifies a direct correspondence between translocation of a DNA molecule through the nanocapillary with the ionic current events detected in typical nanopore experiments. A close-up of the ionic current event is shown as an inset in Figure\refer{figure1}{b}. It also demonstrates that the current blockade corresponds to only a very brief period ($<$1 ms) when the molecule is right at the tip.

\begin{figure}[h!]
\includegraphics {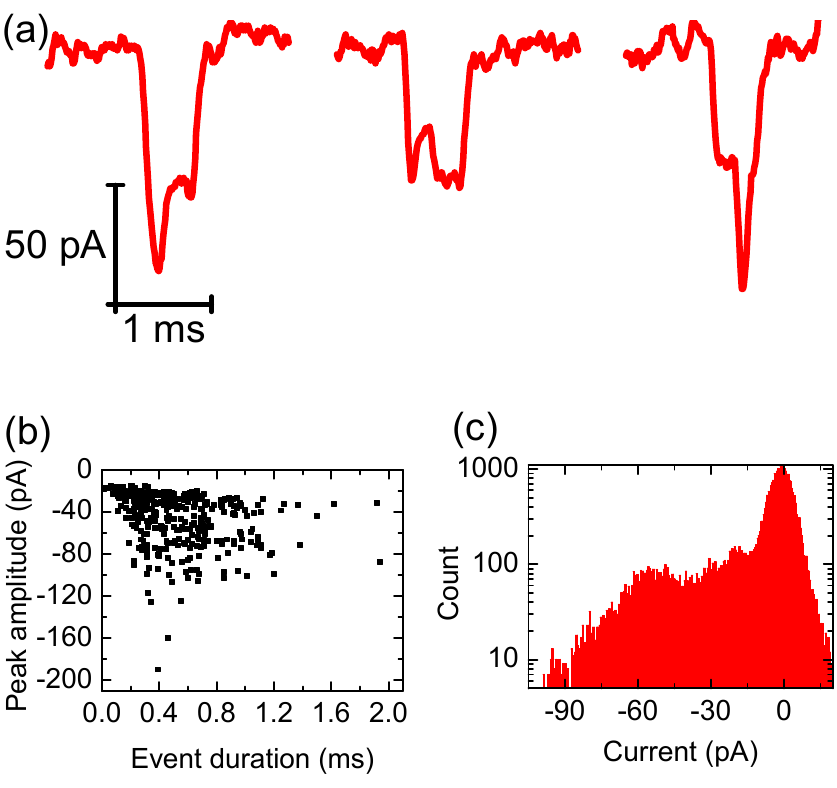}
\caption{\label{fig:figure2}(a) Three typical events showing DNA folding. (b) A scatter plot for 346 indvidual translocation events. (c) A histogram of the translocation events filtered at 5 kHz using a cutoff of 30 pA amplitude and 300 $\mu$s duration to remove degraded DNA events.}
\end{figure}

To determine whether the presence of the intercalating dye molecules was significantly altering the behaviour of the DNA molecules, we detected many ionic current translocation events. Statistics from the ionic current detection at an applied potential of 800 mV is presented in Figure~\ref{fig:figure2}. A feature of translocation of long DNA molecules through nanopores of diameter 10 nm or more is the detection of multiple levels in the ionic current traces due to the DNA molecule folding over itself during translocation. These have also previously been detected with nanocapillaries\cite{Steinbock2010}. In Figure\refer{figure2}{a}, we present three representative translocation events that show two folding levels. This observation is consistent with the mechanical properties of DNA not being altered by the presence of intercalating dye molecules\cite{Gunther2010}. 

An analysis of each translocation event is presented as a scatter plot in Figure\refer{figure2}{b}. The majority of events have a duration between 0.4 ms - 1.2 ms which is consistent with our previous work on nanocapillaries. However, there are some events with mean amplitudes clustered around 20 pA. This is lower than expected and indicates DNA degradation in the presence of the intercalating dye and continuous laser irradiation. We therefore choose to ignore these points by imposing conditions of minimum duration and amplitude on the translocation events. A histogram of these events in Figure\refer{figure2}{c} filtered using this cutoff presents further evidence of the folding levels. There is a broad peak centred at 55 pA indicating DNA degradation and folding. 

We analysed the videos obtained simultaneously with ionic current data presented. Due to the horizontal geometry of the nanocapillaries, the DNA molecules translocate perpendicular to the optical axis. This enables us to track them up to 16  $\mu$m upstream of the pore.  Tracks were obtained for multiple DNA molecules for a range of applied potentials. For the sake of clarity, tracks for just three voltages are presented in Figure\refer{figure3}{a}. The tracking was performed with custom written software in LabVIEW. The schematic in the upper left corner of the figure explains the coordinate system used. We identify $x=0$ with the nanocapillary tip and $t$ with the time elapsed since the start of translocation by the DNA at $x=0$. 

\begin{figure}[h!]
\includegraphics  {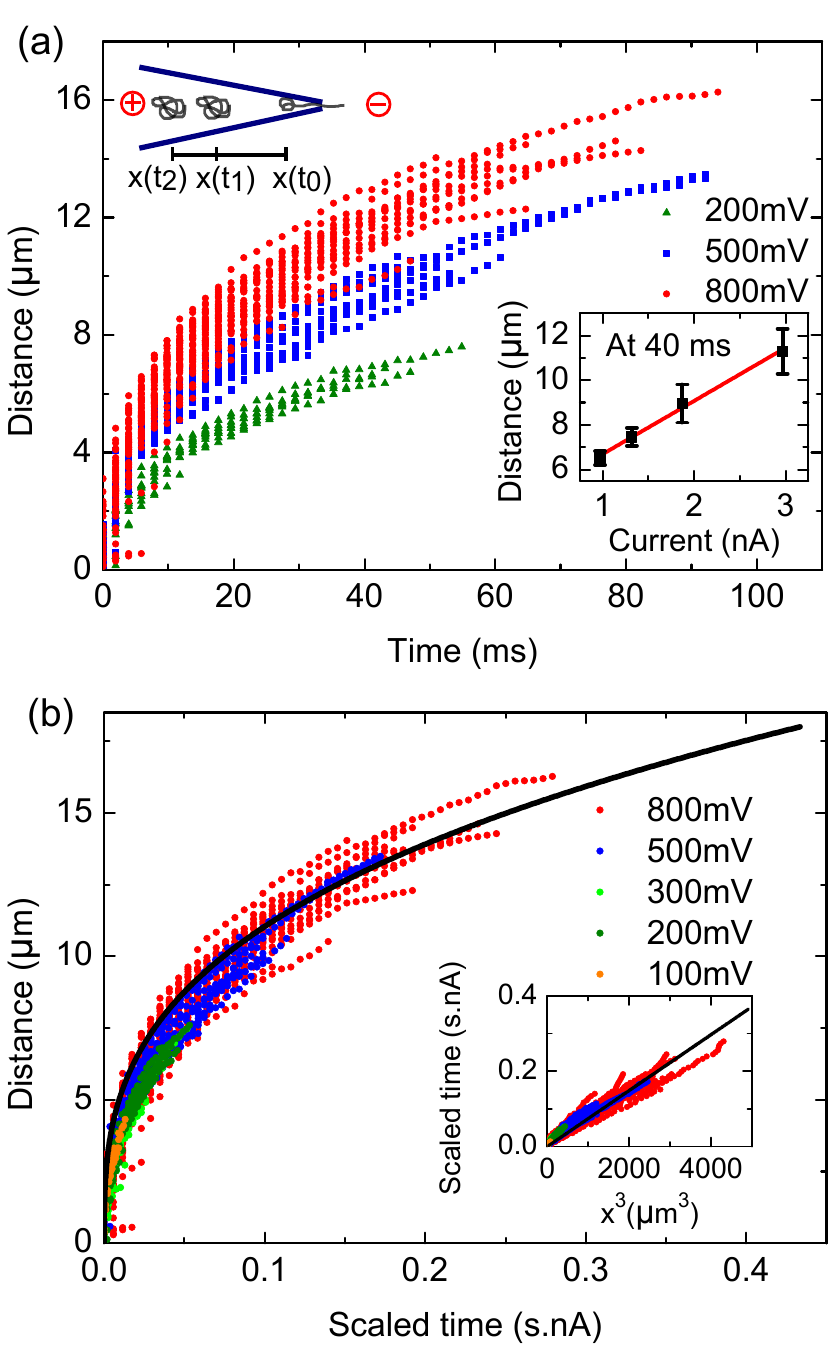}
\caption{\label{fig:figure3}(a) Tracks of multiple DNA molecules for three different voltages. The interval between succesive frames is $\Delta t=1.98$ ms. The distance covered 40 ms from the start of translocation scales linearly with applied voltage (Inset). (b) The tracks for all voltages have been scaled and fitted with a model described in equations~\ref{eqn:eqn4} and~\ref{eqn:eqn5}.}
\end{figure}

It is clear from Figure\refer{figure3}{a} that the DNA slows down as it moves up the nanocapillary. The distance moved in a fixed time depends on the applied voltage. The latter effect is clearly due to the fact that the DNA velocity is proportional to the electric field, which in turn, is proportional to the current. This is easily verified by plotting the distance traveled in a fixed time interval (we choose 40 ms) as a function of the ionic current which gives a linear plot (inset, Figure\refer{figure3}{a}).

The DNA moves due to a combination of electrophoresis and advection by the electroosmotic flow setup in the capillary. Both of these effects are proportional to the electric field\cite{Ghosal2007} $E$. So the equation of motion for the DNA molecules is
\begin{equation}
 \frac{dx}{dt}=\mu_{net} E
\label{eqn:eqn1}
\end{equation}
 where $\mu_{net}$ is the magnitude of the algebraic sum of the electrophoretic and electroosmotic mobilities. From Ohm's law, we have
\begin{equation}
E= \frac{J}{\sigma}=\frac{I}{\sigma A(x)}
\label{eqn:eqn2}
\end{equation}
where $J$ is the ionic current density, $I$ is the ionic current through the pore, $\sigma$ is the conductivity of the buffer solution and the cross section of the pore as a function of distance into the nanocapillary is given by $A(x)$. Integrating Equation~\ref{eqn:eqn1}, we obtain the general relation for time \textit{t} which is valid for a nanocapillary of any geometry
\begin{equation}
t = \frac{\sigma}{\mu_{net} I}\int_0^x\!A(y)\, dy
\label{eqn:eqn3}
\end{equation}

Assuming a conical geometry for the nanocapillary we have $A(x) = \pi r^2 = \pi (a +x\tan\theta)^2$ where $a$ is the pore radius at the capillary tip. Substituting in Equation~\ref{eqn:eqn3}, we get an integral that can be evaluated analytically. We get a particularly simple form if we neglect the pore radius by setting $a=0$. Then the integrand has a singularity at $x=0$ but the singularity is integrable. On evaluating the integral we obtain the simple result $t \propto x^3$. it is convenient to write this result as

\begin{equation}
t' =mx^3, \text{where } t'=t\cdot I
\label{eqn:eqn4}
\end{equation}
which gives us
\begin{equation}
\mu_{net}=\frac{\sigma \pi \tan^2\theta}{3m}
\label{eqn:eqn5}
\end{equation}

The data in Figure\refer{figure3}{a} has been re-plotted as $x$ as a function of $t'$ in Figure\refer{figure3}{b}. The measurements at different voltages are seen to collapse to a single curve as expected. The line is a parameter fit using the form of Equation~\ref{eqn:eqn4}. The inset shows that $t'$ is indeed proportional to $x^3$ as expected with a proportionality constant that is independent of the applied voltage. The conductivity of the buffer is $\sigma=17.93$ mS/cm. The opening angle of the nanocapillaries is determined by optical microscopy to yield a value of  $\tan\theta=0.0187$ and so using equation~\ref{eqn:eqn5} we obtain $\mu_{net} = (8.83\pm 0.17) \times 10^3 \mu$m$^2$/sV. This value is less than the electrophoretic mobility obtained in bulk for $\lambda$ DNA in the same buffer using laser Doppler electrophoresis ($2.6 \times 10^4 \mu$m$^2$/sV). It is also half the value reported for DNA confined between two nanopores\cite{Langecker2011}($1.8\times 10^4 \mu$m$^2$/sV). These results can be explained by a number of other factors including a reduced charge on the DNA backbone due to the presence of the intercalating dye molecules, low pH of the buffer or larger electroosmostic flow out of the nanocapillary. 

In summary, using a setup that combines ionic current detection with fluorescence imaging we present direct visual evidence that current blockade events do indeed correspond to the translocation of DNA through the pore. The current blockade is a sensitive determinant of pore translocation as the drop in the current occurs only when the DNA molecule is in close proximity to the pore. Once across the pore region, the DNA moves in a variable electric field due to the changing cross-section of the nanocapillary. By tracking the motion of the DNA we show that the observed motion is consistent with a simple analytical model. Fitting the single parameter in the model to the data provides a method of measuring the net mobility of the DNA. The measured value was found to be in the same range as other reported measurements of DNA moving in confined environments.

\begin{acknowledgments}
We thank S. Pagliara for help with tracking software in LabVIEW. VVT gratefully acknowledges funding from the Cambridge Commonwealth Trust, the Jawaharlal Nehru Memorial Trust and the Emmy Noether program. SG acknowledges funding from the NIH (USA) under grant R01HG004842 and from the Leverhulme Trust (UK). SHA, NAWB and UFK acknowledge funding from an ERC starting grant, the EPSRC NanoDTC program and the Emmy Noether program, respectively.
\end{acknowledgments}


\end{document}